\documentclass[10 pt,conference, letterpaper]{IEEEtran}

\usepackage{fancyhdr, epsfig, epsf, amsthm, amsmath, amssymb, amsfonts, subfigure, color,soul}
\usepackage[noadjust]{cite}
\usepackage{dsfont}
\usepackage{enumerate}
\usepackage{comment}
\usepackage{bigstrut}

\def\E{\mathsf{E}}

\def\SNR{\mathsf{SNR}}

\def\l{\left}
\def\r{\right}
\def\({\left(}
\def\){\right)}

\def\[{\left[}
\def\]{\right]}

\def\V{\V}
\def\up{\text{up}}
\def\dn{\text{dn}}
\def\bin{b^\up}
\def\bout{b^\dn}

\def\N{\mathbf{N}}
\def\NN{\mathcal{N}}

\def\x{\text{X}}
\def\w{\mathbf{w}}
\def\b{\text{b}}
\def\comp{\text{c}}
\def\V{\mathcal{V}}
\def\B{\mathcal{B}}

\def\ww{\text{w}}

\def\papertitle{Minimizing Latency to Support VR Social Interactions over Wireless Cellular Systems via Bandwidth Allocation}

\IEEEoverridecommandlockouts

\begin{document}
\title{\fontsize{22}{30} \selectfont  \papertitle}

\author{Jihong~Park, $^\dagger$Petar Popovski, and $^\ddagger$Osvaldo~Simeone
\thanks{J.~Park was with Aalborg University, and now he is with Centre for Wireless Communications, University of Oulu, Finland (email: jihong.park@oulu.fi).} 
\thanks{$^\dagger$P.~Popovski is with Department of Electronic Systems, Aalborg University, Denmark (email: petarp@es.aau.dk).}
\thanks{$^\ddagger$O.~Simeone is with Department of Informatics, King's College London, United Kingdom (email: osvaldo.simeone@kcl.ac.uk).}
\thanks{This work has been supported by the European Research Council (ERC Grant Nos. 648382 and 725731) within the Horizon 2020 Program.}
}

\maketitle \thispagestyle{empty}
\begin{abstract} 
Immersive social interactions of mobile users are soon to be enabled within a virtual space, by means of virtual reality (VR) technologies and wireless cellular systems. In a VR mobile social network, the states of all interacting users should be updated synchronously and with low latency via two-way communications with edge computing servers. The resulting end-to-end latency depends on the relationship between the virtual and physical locations of the wireless VR users and of the edge servers. In this work, the problem of analyzing and optimizing the end-to-end latency is investigated for a simple network topology, yielding important insights into the interplay between physical and virtual geometries.
\end{abstract}
\begin{IEEEkeywords} Virtual reality (VR), social network, latency, resource management.
\end{IEEEkeywords}

\section{Introduction}
Virtual reality (VR) is a key use case for 5G \cite{ABIQualcommVR:17,EjderVR:17,MohammedWCNC:18,NokiaVR:18}. Its emergence is powered by the recent advances in computing, which enable immersive real-time interactions with virtual objects. As announced by Facebook \cite{FBspaces:17} and Microsoft \cite{MicrosoftMR}, users will soon be able to interact with each other within virtual communities using VR technologies. In this paper, we consider the problem of supporting \emph{VR-based mobile social networks} over cellular systems by means of edge computing~\cite{ABIQualcommVR:17,EjderVR:17,MohammedWCNC:18,NokiaVR:18}.

A key new element of this challenging problem is the discrepancy between virtual and physical locations of the participating users. In fact, traffic is generated by VR communities in a virtual space, but the supporting network resource for communication and computation are located within the physical network infrastructure. Therefore, the users in the same VR community may not always be co-located in the physical space. For example in Fig.~1, user C belonging to VR community 1 is close in the physical space to user D affiliated to VR community 2, but far from users A and B in VR community 1. This spatial difference between virtual and physical topologies affects the operation of resource allocation and transmission techniques over both Radio Access Network (RAN) and backhaul.

To elaborate, consider VR mobile users that interact in a virtual space. In order for these interactions to be perceived as natural, the network needs to guarantee low latency of e.g. $10$~ms for tactile interactions \cite{ABIQualcommVR:17}. At the same time, all user states should be properly synchronized in the shared virtual environment. Each user's end-to-end latency is thus dominated by the user from the VR community that experiences the worst latency, accrued due to communication and processing. This, in turn, depends on the physical distribution of users belonging to the same VR community and on the spatial availability of communication and computation resources within the physical network infrastructure.

In this work, we study the problem of supporting a VR mobile social network over a multi-cell wireless cellular system with the goal of minimizing the end-to-end latency. Specifically, we focus on the problem of minimizing the end-to-end latency via the bandwidth allocation of the uplink and downlink channels used for communication between users and computing servers. To this end, we formulate a simple model based on a linear cellular topology~that captures the interplay between the social interactions within the VR mobile social network and the location of the computation and communication resources within the physical network. The average end-to-end latency is evaluated by accounting for the contributions of uplink, downlink, and backhaul transmissions, as well as for processing times at the servers. The resulting latency is minimized through a stochastic optimization technique.

\begin{figure}\label{Fig:Network}
	\centering
	\includegraphics[width=8.37 cm]{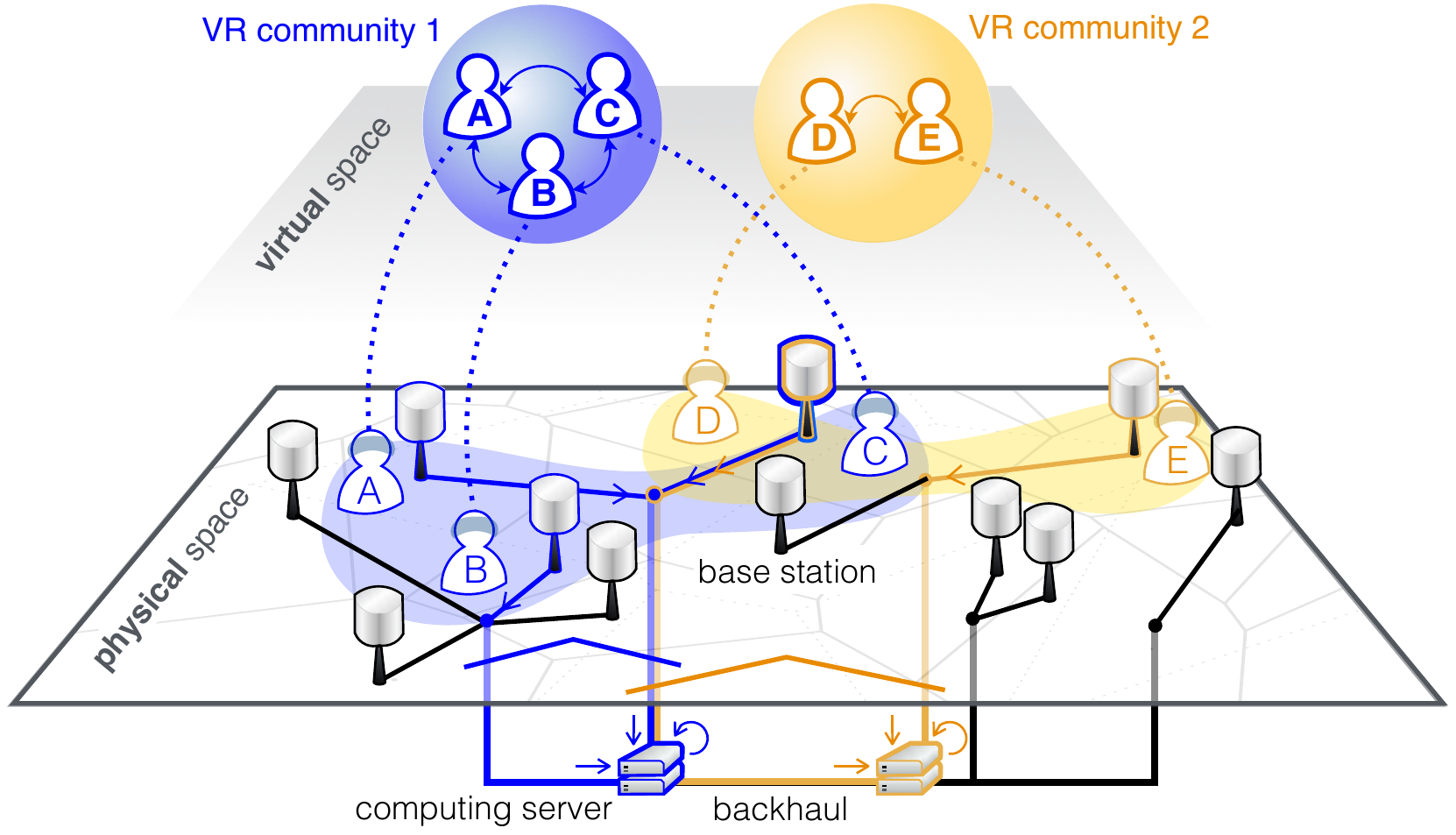}
	\caption{Illustration of VR mobile social network, where the traffic is generated by virtual-space user interactions and supported by a physical cellular network.}
\end{figure}

\textbf{Related Works} --
Current VR headsets provide wireless connections via WiFi and/or WiGig (60 GHz) technologies using unlicensed frequency bands \cite{Displaylink}. The resulting short-range barrier can be overcome by enabling 5G wireless connections. For such 5G-enabled VR headsets, computing tasks will be conceivably offloaded to edge-cloud servers, in order to overcome the restrictions brought by the limited computing capability and battery capacity of mobile devices. 

The required wireless capacity needed to support immersive VR experiences has recently been investigated in~\cite{EjderVR:17}. To minimize the VR traffic volume, a caching approach has been proposed in~\cite{MohammedWCNC:18}. In a VR theater scenario, a multicast design has been studied in~\cite{NokiaVR:18}. These works \cite{EjderVR:17,MohammedWCNC:18,NokiaVR:18} focus~only on optimizing the downlink operations. For augmented reality (AR) applications, the optimization of both uplink and downlink transmissions in terms of end-to-end latency has been studied in~\cite{OsvaldoAR:17} for a single-cell scenario. Due to the focus on AR, end-to-end latency model of reference~\cite{OsvaldoAR:17} does not take into account virtual social interactions. Finally, virtual social interactions underlie Massively Multiplayer Online game (MMO) applications such as Second Life~\cite{SecondLife}. Within more restricted virtual spaces, immersive VR social interactions have been recently provisioned by Facebook and Microsoft~\cite{FBspaces:17,MicrosoftMR}. 

\begin{figure}
\centering
    \includegraphics[width=9 cm]{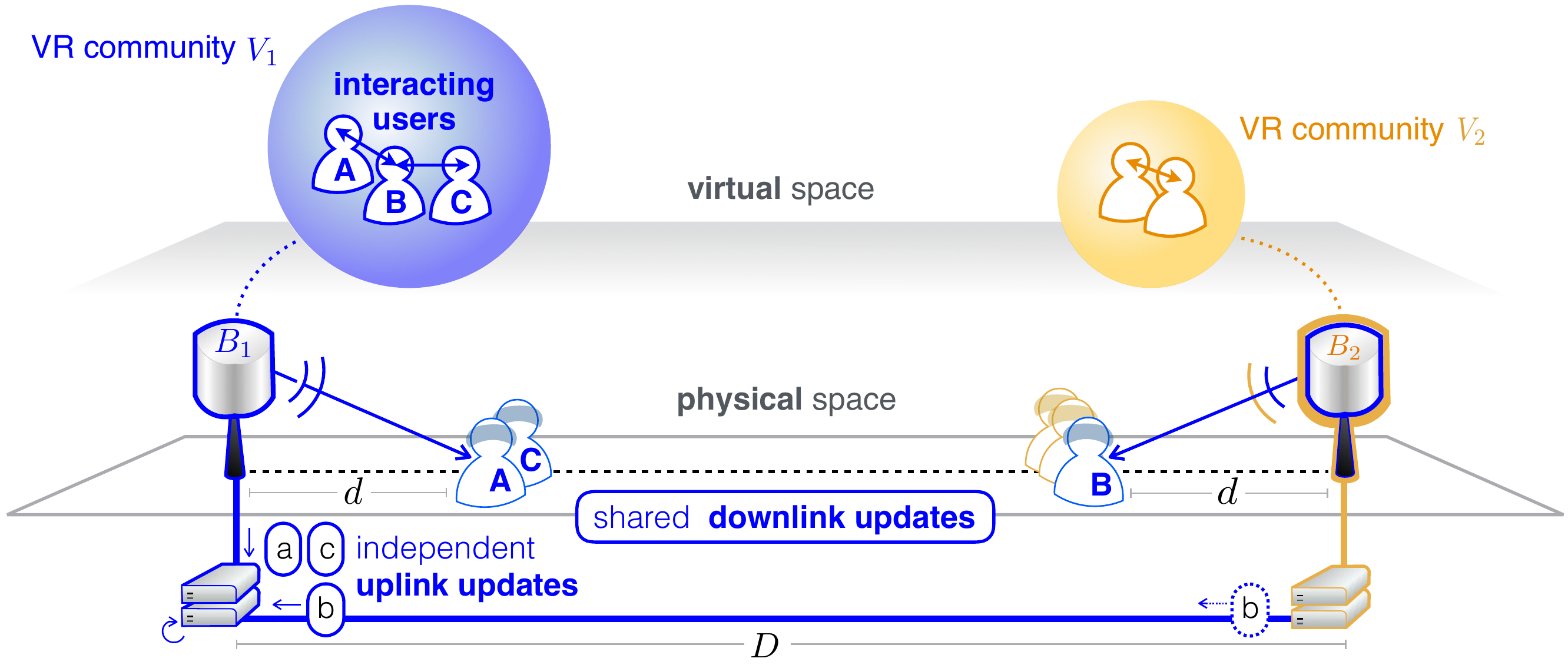}
    \caption{An illustration of a virtual space traffic flows in a one-dimensional physical network model. In the virtual space, VR users A, B, and C interact with each other within VR community $V_1$. Their uplink updates $a$, $b$, and $c$ are sent to the cloud computing server at $B_1$ via \emph{unicast} transmissions. The resulting downlink update needs to be sent to all three users for synchronous interactions via \emph{multicast} transmissions}. 
\end{figure}

\section{System Model}
This section describes the physical network infrastructure, including RAN, backhaul, and computation resources, as well as the VR data traffic model. To focus on the key ideas, we consider a one-dimensional physical network model with two VR communities, as well as two base stations (BSs) in the physical space, as illustrated in Fig.~2. 

We use the subscript $i\in\{1, 2\}$ to indicate VR communities 1 or 2. The subscript $j\in\{1, 2\}$ identifies the two BSs. The superscript $\x \in\{\up,\dn\}$ describes uplink ($\up$) or downlink~($\dn$) operations at a BS.

\subsection{Physical Network and Channel Model} \label{Sect:SysPhy}
The network under study comprises a set of users $\mathcal{K}=\{1, 2, \cdots, N\}$ and two BSs $B_{1}$ and $B_{2}$. Each BS is equipped with a computing server that supports a single VR community. The computing server for VR community $V_i$ is located at BS $B_i$, unless otherwise specified. In the virtual space, each VR community $V_i$ includes a subset of users $\V_i=\{1, 2,\cdots,N_{V_i}\}\subseteq\mathcal{K}$, representing a fraction $p_{V_i}=N_{V_i}/N$ of the set of users, with $p_{V_1}+ p_{V2}=~1$. Furthermore, a subset of users $\B_j = \{1,2,\cdots,N_{B_j}\}\subseteq\mathcal{K}$ associates with BS $B_j$ for both uplink and downlink transmissions in the physical space, representing a fraction $p_{B_j}=N_{B_j}/N$, with $p_{B_1}+p_{B_2}=1$. There are four types of possible user assignments in the virtual and physical spaces, partitioning the set of users into four subsets that are defined as $\NN_{ij}= \mathcal{V}_i \cap \mathcal{B}_j$ with $i\in\{1,2\}$ and $j\in\{1,2\}$. Each type $\NN_{ij}$ includes $N_{ij}\leq\min\{N_{V_i},N_{B_j}\}$ users, representing a fraction $p_{ij}=N_{ij}/N$. Note that we have the equalities $p_{i1} + p_{i2}=p_{V_i}$, and $p_{1j} + p_{2j}=p_{B_j}$.

The two BSs are located at the edges of a one-dimensional physical space with length~$D$, and are connected by a wired backhaul. The BSs use disjoint spectrum bands, hence not interfering with each other. BS $B_j$ assigns orthogonal bands $W^\up$ to uplink and to $W^\dn$ downlink following Frequency Division Duplex (FDD). In the uplink, BS $B_j$ serves each of the $N_{B_j}$ associated users via Frequency Division Multiple Access (FDMA) using {\emph{unicast} transmissions. In the downlink, instead, each BS $B_j$ uses orthogonal \emph{multicast} transmissions in order to update the users in the two VR communities. For a given user configuration $\N=\{N_{11}, N_{12}, N_{21}, N_{22}\}$, we denote by $w_{ij}^\up$ the bandwidth allocated in the uplink to \emph{each user} from the subset $\NN_{ij}$ and by $w_{ij}^\dn$ the bandwidth allocated for multicasting to \emph{all users} in $\NN_{ij}$. The users associated with any BS are located at a distance $d\leq D/2$ as illustrated in Fig.~2. For the purpose of obtaining worst-case performance results, all users are assumed to lie at the maximum distance~$d$. Extensions of the analysis to the more general scenario of arbitrary user-BS distances that are upper bounded by $d$ are possible, but call for more cumbersome notation.

Users in the subsets $\NN_{ij}$ with $j\neq i$, denoted as \emph{cross-type} users, are in VR community $V_i$, but are associated with BS $B_{j}$. The uplink data of these cross-type users must be forwarded through a wired backhaul to BS $B_{i}$ in order to be processed by its attached server. As defined below, each backhaul transmission entails a random delay with the average value proportional to the distance $D$ and the data size. 

For the given physical distance $d$ between a BS and the assigned user, the signal-to-noise ratio ($\SNR$) is determined by path loss attenuation $d^{-\alpha}$ for $\alpha\geq 2$ and by independent Rayleigh~fading. Therefore, the $\SNR$ in uplink or downlink for a given user is

\vspace{-10pt}\small\begin{align}
\SNR^X &= P^X  d^{-\alpha} g/ \sigma^2, \label{Eq:SINR}
\end{align}\normalsize
where $P^X$ denotes transmit power and $\sigma^2$ indicates the noise variance. The term $g$ represents a small-scale fading coefficient that follows an exponential distribution with unitary mean. These coefficients are independently and identically distributed (i.i.d.) across users in uplink and downlink. We assume the use of type-I Hybrid Automatic Repeat reQuest (HARQ), while the instantaneous $\SNR$ information is unknown at the BSs.

\subsection{Virtual Space Traffic} \label{Sect:SysVRtraffic}
In the virtual space, all the $N_i$ VR users in community $V_i$ are assumed to interact with each other, directly or indirectly, as seen in Fig.~2. In order to enable these virtual interactions, each user uploads its uplink state update message with size $\bin$ at regular time intervals, and all the $N_i$ users download the common downlink update message with size $\bout$. We hereafter consider a fixed users' allocation in physical and virtual spaces given by $\N=\{N_{ij}\}$. With this given user configuration $\N$, we focus on a reference user $o$. This user $o$ is uniformly randomly selected in the set of $N$ users, and thus has a type $\NN_{ij}$ with probability $p_{ij}=N_{ij}/N$. For a single state update, the VR traffic of user $o$ is characterized by the following~phases.
\begin{itemize}
\item \textbf{Step 1 (Upload)} -- The user $o\in\NN_{ij}$ uploads its update message with size $\bin$ bits to the associated BS $B_j$. If user $o$ is of cross-type, i.e. $j\neq i$, the uplink data is forwarded to the desired computing server at $B_{i}$ through the inter-BS wired backhaul;

\item\textbf{Step~2 (Compute)} -- The computing server at BS $B_i$ collects all input data from user $o$ as well as from its interacting $N_{V_i}-1$ users, and then produces their synchronous output states;

\item\textbf{Step~3 (Download)} -- The output states are updated with a common message of size $\bout$ bits to all $N_{V_i}$ users through wireless and, for cross-type users, backhaul links.
\end{itemize}

Note that, in order to carry out Step~2, the computing server needs to collect data from all $N_{V_i}$ users in VR community $V_i$. For this reason, the delay prior to computing is limited by the user with \emph{the worst uploading delay in VR community $V_i$}, as described next.

\subsection{Physical Space Delay}

In this section, we fix the user configuration $\N$ and the spectrum allocation $\w=\{\w^\up, \w^\dn\}$ with $\w^\up = \{w_{ij}^\up\}$ and $\w^\dn = \{w_{ij}^\dn\}$ for $i,j\in\{1,2\}$, and analyze the latency of a reference user $o$ with a fixed type $\NN_{ij}$. According to the described VR input/output data flow, conditioned on $\N$, $\w$, and the reference user's type, the average end-to-end latency $T_{ij}(\N,\w)$ of user $o\in\NN_{ij}$ consists of the average uploading delay $T_i^\up(\N,\w^\up)$, computing delay $T_i^{\comp}(\N)$, and downloading delay $T_{ij}^\dn(\w^\dn)$ as in

\vspace{-10pt}\small\begin{align}
T_{ij}(\N,\w) &= T_{i}^\up(\N,\w^\up) + T_{i}^{\comp}(\N) + T_{ij}^\dn(\w^\dn). \label{Eq:E2ELatency}
\end{align}\normalsize

We now discuss the three terms in \eqref{Eq:E2ELatency}. First, the average uploading delay $T_{i}^\up(\N,\w^\up)$ is, as discussed, the worst user's uploading delay for the $N_{V_i}$ users in VR community $V_i$. Denoting~by $D_\ell^{\ww.\up}(\w^\up)$ and $D_\ell^{\b.\up}$ the instantaneous uplink wireless and backhaul delays for any user $\ell\in\V_i$, the average uploading delay~is given as

\vspace{-10pt}\small\begin{align}
\hspace{-5pt}(\textbf{Upload})\;\; T_{i}^\up(\N,\w^\up) = \E\big[ \max_{\ell\in \mathcal{V}_{i}} \big\{ D_{\ell}^{\ww.\up}(\w^\up) + D_\ell^{\b.\up} \big\}\big]. \label{Eq:DelayUpload}
\end{align}\normalsize
In \eqref{Eq:DelayUpload}, the expectation is taken over the random number of transmission time slots required by the HARQ process as well as the random backhaul delay, as detailed next.

The uplink wireless delay $D_{\ell}^{\ww.\up}(\w^\up)$ in \eqref{Eq:DelayUpload} depends on the instantaneous uplink $\SNR$s, which are random due to the small-scale fading coefficients in \eqref{Eq:SINR}. Specifically, if the instantaneous $\SNR^\up$ is no smaller than a target threshold $\theta^\up$, the received signal is successfully decoded; otherwise, a retransmission occurs. For a target success probability~$\eta$, such that $\Pr(\SNR^\up\geq~\theta^\up)=\eta$, the threshold is given as

\vspace{-10pt}\small\begin{align}
\theta^\up=P^{\up}\log_2 \(\eta^{-1}\)/(d^\alpha\sigma^2).
\end{align}\normalsize
As a result, the number $M_{\ell}^\up$ of transmission attempts by the $\ell$-th user follows a geometric distribution with mean $1/\eta$. Measuring the achievable rate via Shannon capacity, each transmission lasts for $\bin/[w_\ell^\up \log_2(1+\theta^\up)]$ seconds, where the uplink spectrum allocation equals $w_\ell^\up=w_{ij}^\up$ if user~$\ell\in\NN_{ij}$. The total uplink wireless transmission delay of the user $\ell$ is hence given~as

\vspace{-10pt}\small\begin{align}
D_{\ell}^{\ww.\up}(\w^\up) &=  M_\ell^\up \bin /\[ w_{\ell}^\up \log_2(1 + \theta^\up) \],
 \label{Eq:LWireless}
\end{align}\normalsize
where $M_\ell^\up \sim \textsf{Geometric}\(\eta \)$.

The uplink backhaul delay $D_\ell^{\b.\up}$ in \eqref{Eq:DelayUpload} is non-zero only for a cross-type user $\ell\in\NN_{ij}$, with $j\neq i$. We define the indicator $\delta_\ell$ such that $\delta_\ell=1$ if user $\ell$ is of cross-type, and otherwise we have $\delta_\ell=0$. Following \cite{Marios:16}, we model the random backhaul delay for cross-type users to follow a Gamma distribution

\vspace{-10pt}\small\begin{align}
D_\ell^{\b.\up} = \delta_\ell B^\up, \label{Eq:LWired}
\end{align}\normalsize
where $B^\up \sim \textsf{Gamma}(D, c_b \bin )$. The constant $c_b>0$ represents the propagation delay per bit and unit backhaul length, e.g., $c_b=10^{-8}$ seconds \cite{Marios:16}. Note that the average backhaul delay of cross-type users is $c_b D b^\up$, and is hence proportional to $D$ and $b^\up$.

Second, the average computing delay $T_i^{\comp}(\N)$ of user $o\in\NN_{ij}$ in \eqref{Eq:E2ELatency} is the server processing time required to update the VR model based on input data from the $N_{V_i}$ users in community $V_{i}$. With the clock speed $f_s$, the processing of the VR input data of size $\bin$ per user requires $\bin/f_s$ seconds. The average computing delay is thereby given as:

\vspace{-10pt}\small\begin{align}
\hspace{-20pt}(\textbf{Compute})\quad T_{i}^\comp(\N) = \bin N_i/f_s. \label{Eq:DelayCompute}
\end{align}\normalsize

Finally, the average downloading delay $T_{ij}^\dn(\w^\dn)$ of user $o$ in \eqref{Eq:E2ELatency} comprises the average downlink wireless transmission delay induced by the HARQ process and the wired backhaul delay. Denoting as $D^{\ww.\dn}(\w^\dn)$ and $D^{\b.\dn}$ the instantaneous downlink wireless transmission and backhaul delays, the average downloading delay for user $o$ is computed as

\vspace{-10pt}\small\begin{align} 
\hspace{-20pt} (\textbf{Download})\quad T_{ij}^\dn(\w^\dn) = \E[D^{\ww.\dn}(\w^\dn)] + \E[D^{\b.\dn}],  \label{Eq:DelayDownload}
\end{align}\normalsize
where the two terms are discussed next.

For downlink transmissions, BS $B_j$ keeps multicasting to the associated $N_{ij}$ users until all of them successfully decode the output update message. Nevertheless, the latency for user $o\in\NN_{ij}$ depends solely on its own decoding process. Following the same reasoning as for \eqref{Eq:LWireless}, the number $M^\dn$ of transmission attempts follows a Geometric distribution, resulting in the average downlink wireless transmission delay

\vspace{-10pt}\small\begin{align}
\E[D^{\ww.\dn}(\w^\dn)] &= b^\dn /[w_{ij}^\dn \eta \log_2(1+ \theta_{ij}^\dn)]. \label{Eq:DLwirelessDelay}
\end{align}\normalsize
The downlink $\SNR$ threshold $\theta_{ij}^\dn$ in \eqref{Eq:DLwirelessDelay} is set such that all $N_{ij}$ users can successfully decode their received multicast signals with probability $\eta$. Denoting $\SNR_\ell$ as user $\ell$'s $\SNR$, this implies the condition $\Pr(\min_{\ell\in\NN_{ij}}\{\SNR_\ell\}\geq \theta_{ij}^\dn)=\eta$, which yields

\vspace{-10pt}\small \begin{align}
\theta_{ij}^\dn = P^{\dn}\log_2 \(\eta^{-1}\)/(d^\alpha\sigma^2 N_{ij}).
\end{align}\normalsize

For the backhaul term, as in \eqref{Eq:LWired}, a backhaul transfer occurs when the user $o$ is of cross-type. This entails a random delay following $\textsf{Gamma}(D,c_b b^\dn)$ distribution, yielding the average downlink backhaul delay

\vspace{-10pt}\small\begin{align}
\E[D^{\b.\dn}]&=\delta_o c_b  b^\dn D, \label{Eq:DLbackhaulDelay}
\end{align}\normalsize
where $\delta_o=1$ if user $o$ is of cross-type, and $\delta_o=0$~otherwise.

\section{ End-to-End Latency Minimization}
The objective of this section is to tackle the minimization of the average end-to-end latency studied in the previous section with respect to the spectrum allocation $\w$ for a given user configuration $\N$. Besides the average taken in \eqref{Eq:E2ELatency} over HARQ and backhaul delays, here we further consider the expectation over the choice of a reference user $o$. Assuming a uniformly distributed selection, user $o$ belongs to the type $\NN_{ij}$~with probability $p_{ij}=N_{ij}/N$, in which case it experiences the average end-to-end latency $T_{ij}(\N,\w)$ in \eqref{Eq:E2ELatency}. The optimal spectrum allocation $\w^* = \{\w^{\dn^*},\w^{\up^*}\}$~minimizing the average end-to-end latency of user~$o$ is then given as

\vspace{-10pt}\small\begin{align}
\w^*= \underset{\l\{\w^\dn, \w^\up\r\}}{\arg\min}\;\;& \sum_{j=1}^2\sum_{i=1}^2  p_{ij} T_{ij}(\N,\w)  \label{Eq:Objective}\\
\text{s.t.}\quad\quad  \sum_{i=1}^2 w_{ij}^\dn &= W^\dn,\; w_{ij}^\dn\geq 0 \quad\forall j\in\{1, 2\} \label{Eq:P1ConstDn}\\
\sum_{i=1}^2 w_{ij}^\up N_{ij} &= W^\up,\; w_{ij}^\up\geq 0 \quad\forall j\in\{1, 2\}. \label{Eq:P1ConstUp}
\end{align}\normalsize
The constraints \eqref{Eq:P1ConstDn} and \eqref{Eq:P1ConstUp} impose that each BS entirely allocates its available bandwidth for downlink multicast and uplink unicast transmissions, respectively. 

Due to the additive form of the end-to-end latency \eqref{Eq:E2ELatency} and to the distinct constraints \eqref{Eq:P1ConstDn} and \eqref{Eq:P1ConstUp}, the optimal allocation $\w^*$ can be achieved by minimizing \eqref{Eq:Objective} separately with respect to the downlink allocation $\w^\dn$ and the uplink allocation $\w^\up$.

For the downlink, using \eqref{Eq:DelayDownload}, one needs to equivalently minimize the objective function $N_{1j}/w_{1j}^\dn + N_{2j}/w_{2j}^\dn$ under the constraint \eqref{Eq:P1ConstDn}. This convex problem can be solved by applying the Karush-Kuhn-Tucker (KKT) conditions, yielding the optimal downlink allocation

\vspace{-10pt}\small\begin{align}
w_{ij}^{\dn^*} = W^\dn \sqrt{N_{ij}}/(\sqrt{N_{1j}}+\sqrt{N_{2j}}). \label{Eq:DnAlloc}
\end{align}\normalsize

For uplink spectrum allocation, using \eqref{Eq:DelayUpload}, the said problem is equivalent to minimizing the objective function $\sum_{i=1}^2 N_{V_i} \E[ \max_{\ell\in \mathcal{V}_{i}}  \{ a M_\ell^\up/w_\ell^\up + D_\ell^{\b.\up} \} ]$ under the constraint \eqref{Eq:P1ConstUp}, with $a=\bin /\log_2(1 + \theta^\up)$. This is a convex but non-differentiable stochastic problem. It can be tackled by means of the stochastic approximation method, whereby the objective function is estimated via its empirical mean with a number $T$ of independent samples of the relevant random variables \cite{Shai:SO}, as further detailed below. The resulting problem can be solved by applying the projected subgradient method with some number $K$ of iterations \cite{Boyd:Subgradient}. Convergence to the global optimum $\w^{\up^*}$ is guaranteed for sufficiently large $T$ and $K$ under suitable technical conditions \cite{Shai:SO,Boyd:Subgradient}.

To elaborate, we draw $T$ independent samples of the transmission attempt numbers \small{$\{M_{\ell,t}^\up\}$}\normalsize\; and of the backhaul delays \small{$\{D_{\ell,t}^{\b.\up}\}$\normalsize, and consider the empirical objective function {\small$1/T\sum_{t=1}^T \sum_{i=1}^2 N_{V_i} \max_{\ell\in \mathcal{V}_{i}}  \{ a M_{\ell,t}^\up/w_\ell^\up + D_{\ell,t}^{\b.\up} \}$}\normalsize. For~a~given spectrum allocation \small{$w_{\ell}^{\up(k)}\in\{w_{ij}^{\up(k)}\}$}\normalsize\; at the $k$-th iteration of the subgradient method, we define the index {\small$\ell_{i,t}^{(k)}$}\normalsize of the user inducing the worst uploading delay in VR community $V_i$ for the $t$-th sample as \small{$\ell_{i,t}^{(k)}=\arg\max_{\ell\in \mathcal{V}_{i}}  \{ a M_{\ell,t}^\up/w_{\ell }^{\up(k)} + D_{\ell,t}^{\b.\up} \}$}\normalsize. The subgradient \small{$g_{ij}^{(k)}$}\normalsize and the uplink allocation \small{$\tilde{w}_{ij}^{\up(k+1)}$}\normalsize\; for the next $(k+1)$-th iteration, prior to the enforcement of the constraint \eqref{Eq:P1ConstUp}, are then given as

\vspace{-10pt}\small\begin{align}
&g_{ij}^{(k)}=- \frac{a N_{V_i}}{T ({w_{ij}^{\up(k)}})^2} \sum_{t=1}^T \delta_{ij}^{(k)} M_{\ell_{i,t}^{(k)},t}^\up\quad \text{and}\\
&\tilde{w}_{ij}^{\up(k+1)} =w_{ij}^{\up(k)}-\beta g^{(k)}, \label{Eq:preWup}
\end{align}\normalsize 
respectively with a step size $\beta>0$. The indicator function equals \small{$\delta_{ij}^{(k)}=1$}\normalsize\; if \small{$\ell_{i,t}^{(k)}\in \NN_{ij}$}\normalsize, and \small{$\delta_{ij}^{(k)}=0$}\normalsize, otherwise. The next-iterate uplink allocation \small{$\{{w}_{ij}^{\up(k+1)}\}$}\normalsize\; is finally obtained by projecting \small{$\{\tilde{w}_{1j}^{\up(k+1)},\tilde{w}_{2j}^{\up(k+1)}\}$}\normalsize\; in \eqref{Eq:preWup} onto the segment~\eqref{Eq:P1ConstUp}. This can be easily seen to yield as follows: \emph{(i)}~if \small{$\tilde{w}_{1j}^{\up(k+1)}<$ $(\tilde{w}_{2j}^{\up(k+1)}-{W^\up}/{N_{2j}}){N_{1j}}/{N_{2j}}$}\normalsize, then \small{$w_{1j}^{\up(k+1)}=0$}\normalsize\; and~\small{$w_{2j}^{\up(k+1)}={W^\up}/{N_{2j}}$}\normalsize; \emph{(ii)} if {\small$\tilde{w}_{2j}^{\up(k+1)}<$}\normalsize\; {\small$(\tilde{w}_{1j}^{\up(k+1)}-{W^\up}/{N_{1j}}){N_{2j}}/{N_{1j}}$}\normalsize, then \small{$w_{1j}^{\up(k+1)}={W^\up}/{N_{1j}}$}\normalsize\; and \small{$w_{2j}^{\up(k+1)}=0$}\normalsize; and \emph{(iii)} otherwise, 

\vspace{-10pt}\small\begin{align}
w_{1j}^{\up(k+1)} &= \frac{ {N_{1j}} W^\up - {N_{1j}}{N_{2j}} \tilde{w}_{2j}^{\up(k+1)} + ({N_{2j}})^2 \tilde{w}_{1j}^{\up(k+1)}  }{({N_{1j}})^2 + ({N_{2j}})^2 },\\
w_{2j}^{\up(k+1)} &= \frac{ {N_{2j}} W^\up - {N_{1j}}{N_{2j}} \tilde{w}_{1j}^{\up(k+1)} + ({N_{1j}})^2 \tilde{w}_{2j}^{\up(k+1)}  }{({N_{1j}})^2 + ({N_{2j}})^2 }.
\end{align}\normalsize

In the next section, the optimal allocation scheme is compared with the equal allocation baselines for downlink multicast and uplink unicast, which are given as

\vspace{-10pt}\small\begin{align}
\hspace{-7pt}(\textsf{Equal Dn})\; w_{ij}^\dn = W^\dn/2 \;\;\text{and}\;\; (\textsf{Equal Up})\;w_{ij}^\up = W^\up/N_{B_j}. \label{Eq:EqualAlloc}
\end{align}\normalsize

\section{Numerical Results and Discussion}
In this section, we evaluate the end-to-end latency of a VR mobile social network under different user configurations in virtual and physical spaces. The virtual and physical geometries determine the number of cross-type users. To capture this important parameter, we define $0\leq \rho_c\leq 1$ as the ratio of cross-type users. Specifically, we consider the symmetric setting with equal loads of the two BSs and with an equal fraction of cross-type users for each BS, i.e. $p_{12}=p_{21}=\rho_c/2$ and $p_{11}=p_{22}=(1-\rho_c)/2$, which satisfies $p_{V_i}=p_{B_j}=0.5$. Other simulation parameters are given as: $N=50$, $b_i^\x = 1$~kbit, $c_b=10^{-8}$ seconds, $f_s=2$~GHz, $W^\x= 1$~GHz, $\eta=0.7$, $P^\up/\sigma^2=20$~dB, $P^\dn/\sigma^2=30$~dB, $\alpha=3$, $D=500$ m, and $d=15$~m. For the proposed stochastic optimization method, we set $T=K=50$.

Fig.~3 shows that the average end-to-end latency versus the cross-type user ratio $\rho_c$. The latency is seen to increase monotonically with $\rho_c$, owing to the increasing backhaul delays of cross-type users. The proposed optimal uplink spectrum allocation is able to partially compensate for the backhaul delays by providing more spectrum to cross-type users. As seen in the figure, this yields up to $25.1$\% latency reduction as compared to the equal allocation in the uplink for the intermediate value of $\rho_c$. Note that optimal uplink spectrum allocation cannot improve the latency performance when we have $\rho_c=0$, i.e., no cross-type users, or $\rho_c = 1$, i.e., all cross-type users. For such cases, all uploading user delays are identically distributed, and it is thus not possible to prioritize spectrum allocation to any group of users. 

We also observe that the mentioned gains are achieved by and large even with equal downlink bandwidth allocation, since the end-to-end latency is mostly dictated by the worst uploading delay. This is due to the fact that, thanks to multicasting and to the larger transmission power of the BSs, downlink transmission delays are typically shorter than unicast uplink transmission delays.

Finally, for $\rho_c>0.5$, it is beneficial to support VR community $V_i$ on the computing server at BS $B_j$ with $j\neq i$, as seen via the dotted curves in Fig.~3. This is obtained by swapping the two VR communities $V_1$ and $V_2$. The said observation emphasizes the importance of the computing server locations, which is an interesting topic for further research. Another possible extension of this work is to incorporate stochastic VR interactions that depend on the virtual-space user locations.

\begin{figure}\centering
\hspace{-5pt}
\includegraphics[width=9cm]{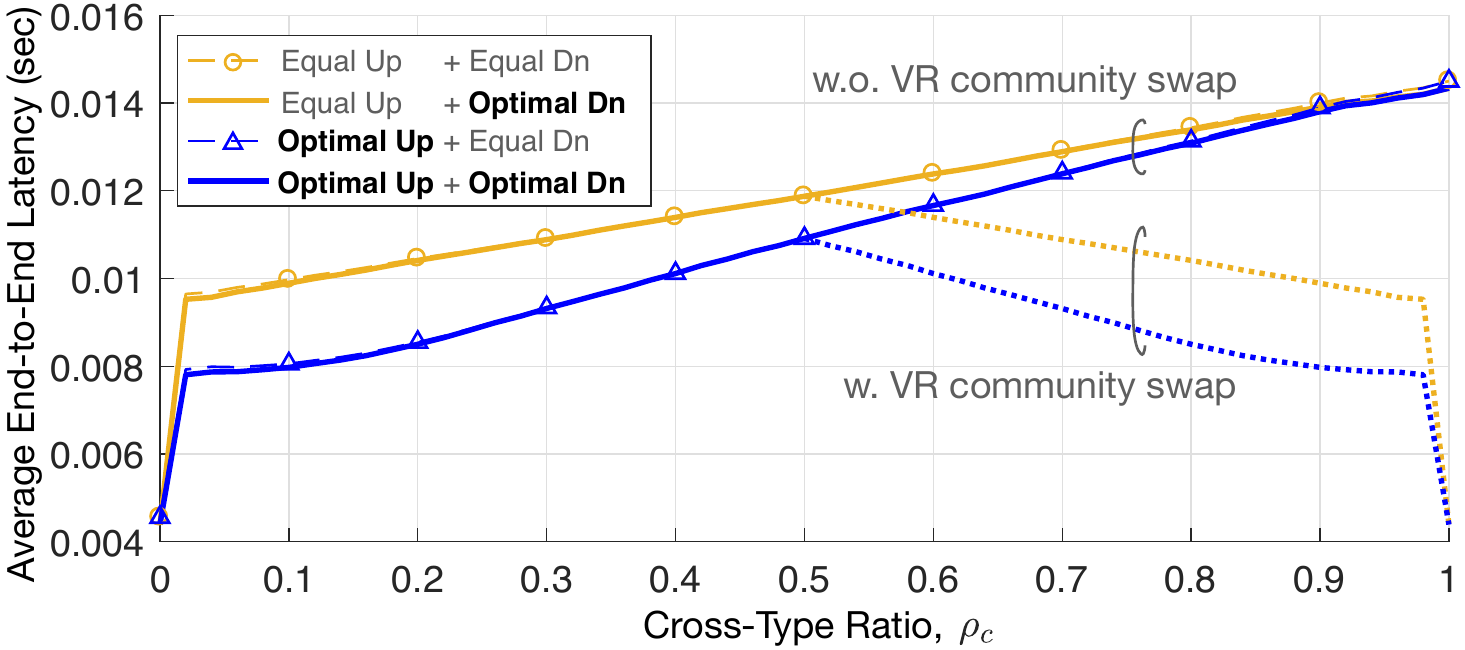}
\caption{Average end-to-end latency of a uniformly randomly selected user~$o$ as cross-type ratio $\rho_c$ increases, where cross-type users connect to the BS that does not run the corresponding VR community. }
\end{figure}

\bibliographystyle{ieeetr}  

\begin{thebibliography}{10}

\bibitem{ABIQualcommVR:17}
{ABI Research and Qualcomm}, ``{Augmented and Virtual Reality: The First Wave
  of 5G Killer Apps},'' {\em White Paper}, Feb. 2017.

\bibitem{EjderVR:17}
{E. Ba{\c s}tu{\u g}, M. Bennis, M. M{\'e}dard, and M. Debbah}, ``{Towards
  Interconnected Virtual Reality: Opportunities, Challenges and Enablers},''
  {\em IEEE Commun. Mag.}, vol.~55, pp.~110--117, Jun. 2017.

\bibitem{MohammedWCNC:18}
M.~S. Elbamby, C.~Perfecto, M.~Bennis, and K.~Doppler, ``{Edge Computing Meets
  Millimeter-wave Enabled VR: Paving the Way to Cutting the Cord},'' {\em to
  appear in Proc. IEEE WCNC 2018.}

\bibitem{NokiaVR:18}
{Athul Prasad, Mikko A. Uusitalo, David Navr{\'a}til, and Mikko S{\"a}ily},
  ``{Challenges for Enabling Virtual Reality Broadcast Using 5G Small Cell
  Network},'' {\em to appear in Proc. IEEE WCNC Wksp. CmMmW5G 2018}.

\bibitem{FBspaces:17}
{Facebook Spaces,} {\em Webpage [Online]. URL: http://facebook.com/spaces}.

\bibitem{MicrosoftMR}
{AltspaceVR}, ``{AltspaceVR joins Microsoft},'' {\em [Online]. URL:
  http://altvr.com/joining-microsoft}.

\bibitem{Displaylink}
{Displaylink,} {\em Webpage [Online]. URL: http://www.displaylink.com/vr}.

\bibitem{OsvaldoAR:17}
A.~AL-Shuwaili and O.~Simeone, ``{Energy-Efficient Resource Allocation for
  Mobile Edge Computing-Based Augmented Reality Applications},'' {\em IEEE
  Wireless Commun. Lett.}, vol.~6, pp.~398--401, Apr. 2017.

\bibitem{SecondLife}
{Second Life,} {\em Webpage [Online]. URL: http://secondlife.com}.

\bibitem{Marios:16}
G.~Zhang, T.~Q.~S. Quek, M.~Kountouris, A.~Huang, and H.~Shan, ``{Fundamentals
  of Heterogeneous Backhaul Design Analysis and Optimization},'' {\em IEEE
  Trans. Commun.}, vol.~64, pp.~876--889, Feb. 2016.

\bibitem{Shai:SO}
S.~Shalev-Shwartz, O.~Shamir, N.~Srebro, and K.~Sridharan, ``{Stochastic Convex
  Optimization},'' {\em Proc. COLT}, 2009.

\bibitem{Boyd:Subgradient}
S.~Boyd, ``{Subgradient Methods},'' {\em notes for EE394b, Stanford
  University}, Spring 2013-2014.

\end{thebibliography}

\end{document}